\documentclass[prd,aps,preprint,showpacs]{revtex4}
\usepackage{epsfig}

\begin{document}


\voffset 1.5cm
\preprint{KIAS-P02076, SNUTP 02-035, hep-ph/0211310}

\title{
\Large 
Bi-large neutrino mixing from bilinear R-parity violation
with non-universality 
}

\author{
 \vspace{2ex}
  Eung Jin Chun$^1$, Dong-Won Jung$^2$ and Jong Dae Park$^2$}

\affiliation{ 
\vspace{1ex}
 $^1$Korea Institute for Advanced Study, P.O.Box 201 Cheongryangri,
 Seoul 130-650, Korea\\
 $^2$School of Physics,
 Seoul National University,
 Seoul 151-747, Korea
\vspace{2cm}  }


\begin{abstract} 
We investigate how the bi-large mixing required by the recent 
neutrino data can be accommodated in the supersymmetric standard model 
allowing bilinear R-parity violation and non-universal soft terms.  
In this scheme, the tree-level contribution and
the so-called Grossman-Haber one-loop diagrams
are two major sources of the neutrino mass matrix.
The relative size of these two contributions falls into the right range
to generate the atmospheric and solar neutrino mass hierarchy.
On the other hand, the bi-large mixing is typically obtained by
a mild tuning of input parameters to arrange a partial 
cancellation among various contributions.
\end{abstract}


\maketitle



\voffset 0cm

Recently, impressive progress has been made in atmospheric and 
solar neutrino experiments \cite{SK,SNO}.  They provided us 
convincing evidences for three active neutrino oscillations requiring
two large and one small mixing angles \cite{CHZ}.
The resulting neutrino mixing matrix \cite{MNS} takes the form;
\begin{equation} \label{mns}
 U \approx \pmatrix{ c_{12} & s_{12} & \theta_{13} \cr
 -{s_{12} \over \sqrt{2}} & {c_{12} \over \sqrt{2}} & {1 \over \sqrt{2}} \cr
 {s_{12} \over \sqrt{2}} & -{c_{12} \over \sqrt{2}} & {1 \over \sqrt{2}} \cr}
\end{equation}
where $c_{ij}=\cos\theta_{ij}$,  $s_{ij}=\sin\theta_{ij}$, and 
$s_{13} \approx \theta_{13} \lesssim 0.2$.
Here we put $\theta_{23}=\pi/4$ for the nearly maximal 
atmospheric neutrino mixing angle.   
The solar neutrino mixing angle $\theta_{12}$ takes the value
$\tan\theta_{12} \approx 0.65$ for the so-called LMA solution which is
strongly favored by the recent SNO data \cite{SNO}.  
The mass-squared differences explaining 
the atmospheric and solar neutrino data are $\Delta m^2_{atm} \approx
2.5\times10^{-3}$ eV$^2$ and $\Delta m^2_{sol} \approx 5\times10^{-5}$
eV$^2$, respectively.  Even though less favored, the so-called LOW solution 
with $\tan\theta_{12} \approx 0.77$ and $\Delta m^2_{sol} \sim 10^{-7}$ 
eV$^2$ is still viable.

\medskip

One of attractive schemes to generate neutrino masses and mixing is
to invoke R-parity and  lepton-number violation allowed in the
supersymmetric standard model \cite{HS}.
The purpose of this paper is to address the question 
whether  the bi-large mixing of three active  neutrinos can 
arise naturally from the bilinear R-parity violation. 
The superpotential of the supersymmetric standard model may contain 
the following bilinear terms;
\begin{equation}
 W = \epsilon_i \mu L_i H_2 \,,
\end{equation}
generalizing the usual $\mu$-term, $\mu H_1 H_2$.  
Then, there are also six soft supersymmetry
breaking terms in the scalar potential;
\begin{equation} \label{Vsoft}
 V_{soft} = \epsilon_i \mu B_i L_i H_2 + m^2_{L_i H_1} L_i H_1^\dagger 
  + h.c. \,,
\end{equation}
where we used the same notations for the superfields and their scalar
components.  Let us note that $B_i$ in the first term is dimension-one 
and the  corresponding term for the Higgs biliear is
$\mu B H_1 H_2$. 

If the universal boundary condition is imposed on the soft-terms, the 
differences between the soft-terms of 
the Higgs boson $H_1$ and slepton $L_i$ such as 
$$\Delta B_i \equiv B-B_i\quad\mbox{and}\quad 
   \Delta m^2_i \equiv m^2_{H_1}-m^2_{L_i},  $$
vanish at the mediation scale of supersymmetry breaking and their non-zero 
values  are generated at the weak scale through renormalization group 
evolution (RGE), while $m^2_{L_i H_1}$ remain vanishing.
In this case, there are only three free parameters $\epsilon_i$ which 
makes the model very economic.  However, this model cannot accommodate 
the bi-large mixing consistently with small $U_{e3}$.
It is easy to understand it qualitatively as one can expect 
that the three parameters $\epsilon_i$ 
control all the mixing angles.  A small $\theta_{13}$ and a large
$\theta_{23}$ requires  $\epsilon_1 \ll \epsilon_2 \approx \epsilon_3$ 
leading to $\theta_{12} \approx \theta_{13}$ \cite{CK,cjkp}. 
Thus, in order to accommodate the bi-large neutrino mixing, one has to
go beyond this minimal scheme.  One way is to allow trilinear 
couplings while keeping the universality.  In this case, the five couplings
related to the third generation fermions may play a major role to 
generate the desired neutrino mass matrix \cite{CK,cjkp}.  
Another way is to allow non-universal soft-terms
\cite{valle1,sacha1,anjan1,abada}.
Introduction of general flavor-mixing soft-masses is, of course, tightly 
constrained by the flavor changing neutral current processes, such as
$\mu \to e \gamma$ or $\tau\to \mu \gamma$ \cite{fcnc}.  
However, the non-universality in the flavor-diagonal soft-parameters
is not severely  constrained.  Generically, one could expect 
$\Delta m^2_i/m_{H_1}^2$ and $\Delta B_i/B$ to be of order one.
One can also have $m^2_{L_i H_1} \sim \epsilon_i m_{H_1}^2$.

In this paper,  we investigate how the desired 
neutrino mass and mixing pattern can arise under such a generic 
non-universality condition.  We will see that the right values 
of the mixing angles and the mass hierarchy can be obtained
in reasonable ranges of parameter space without severe fine-tunning.
In the below, we will first quantify all the tree-level and one-loop
contributions to the neutrino mass matrix  and identify the
dominant contributions.   Obtaining a rather simple form of the 
leading neutrino mass matrix, we will make  qualitative discussions
to understand how the desired masses and mixing arise.
This will be completed by presenting our numerical analysis.

\medskip

Let us start our main discussion by describing the structure of neutrino mass 
matrix coming from R-parity violation.
Adopting the notations of Ref.~\cite{CK}, the most general one-loop
renormalized neutrino mass matrix can be written as
\begin{equation} \label{Mn}
 M^\nu_{ij} = - {M_Z^2 \over F_N} \xi_i \xi_j c_\beta^2 
  - {M_Z^2 \over F_N} (\xi_i \delta_j + \delta_i \xi_j) c_\beta
  + \Pi^\nu_{ij} \, , 
\end{equation}
where $F_N\equiv M_1M_2/M_{\tilde{\gamma}} + M_Z^2 c_{2\beta}/\mu$ with
$M_{\tilde{\gamma}} \equiv c_W^2M_1 + s_W^2M_2$.
Here, the first term is the neutrino mass matrix arising
at tree-level, the second terms
containing $\delta_i$ come from  the one-loop correction to the
neutrino--neutralino mixing masses projected on to the neutrino direction,
and the last term $\Pi^\nu_{ij}$ is the one-loop correction to 
the $\nu_i$-$\nu_j$ Majorana mass matrix.  
The non-zero values of  $\xi_i \equiv \langle L_i^0 \rangle/
\langle H_1^0 \rangle -\epsilon_i$ arise due to non-universal soft terms
in the slepton--Higgs sector as follows;
\begin{equation} \label{xis}
 \xi_i =  \epsilon_i { \Delta m^2_i + \Delta B_i \mu t_\beta \over 
      m^2_{\tilde{\nu}_i} } - {m^2_{L_i H_1} \over m^2_{\tilde{\nu}_i} },
\end{equation}
where the sneutrino mass-squared is 
$m^2_{\tilde{\nu}_i} = m_{L_i}^2 +M_Z^2c_{2\beta}/2$.
As is well-known,  the tree-level mass matrix makes massive 
only one neutrino in the direction of $\vec{\xi}$, which 
is typically the heaviest one, $\nu_3$.  
In fact, the quantity $\xi_i$ controls the neutrino--neutralino mixing
and thus could be probed by lepton flavor violating decays of the lightest
neutralino in the future colliders \cite{jaja,cjkp,stau}.  
Here, let us introduce another quantity,
\begin{equation} \label{etas}
\eta_i \equiv {\langle L_i^0 \rangle \over \langle H_1^0 \rangle} -
\epsilon_i {B_i \over B} = \xi_i +\epsilon_i {\Delta B_i\over B} 
\end{equation}
which governs the mixing between the sleptons and Higgs bosons. 
As we will see, the flavor structure of the neutrino  mass matrix
depends on these two R-parity violating parameters, $\xi_i$ and $\eta_i$, 
as well as non-universal slepton masses.

A simplication of the full neutrino mass matirx comes from the
obseravtion that 
{\it the second term on the right-hand side of Eq.~(\ref{Mn}) 
can be ignored} in our case \cite{sacha2}.  
This can be seen immediately by going to the basis 
where the tree-level mass matrix is diagonalized by the eigenvector 
$\hat{\xi}$ and any two orthogonal unit vectors.  In this basis, 
one finds that the second mass matrix has vanishing components in
the 1-2 plane orthogonal to $\hat{\xi}$.  Thus, leaving the heaviest 
$\nu_3$ untouched, approximate see-saw diagonalization can be applied 
to get the contribution to the 1-2 plane of the order of
$M_Z \delta^2$.  This is like a two-loop contribution much smaller 
than the (non-vanishing) 1-2 components of the last term $\Pi^\nu$.  
Thus, there is no need to compute the second mass term 
in most cases even though we included it in our analysis.

The main contribution to the last term $\Pi^\nu$ of Eq.~(\ref{Mn})
comes from the one-loop diagrams exchanging sneutrinos/Higgs bosons 
and gauginos \cite{GH,sacha2} in the case of generic non-universality 
under consideration.
Here we present the explicit formula of this one-loop mass matrix which
is calculated by the use of approximate see-saw rotation \cite{CK};
\begin{eqnarray} \label{Mgloop}
 \Pi^\nu_{ij} &=& - {g^2\over 32\pi^2} \sum_a (t_W N_{1a}-N_{2a})^2 
                  m_{\tilde{\chi}^0_a}   
  \left( \sum_\phi {1\over2} \theta_{i\phi} \theta_{j\phi} 
         B_0(m_{\tilde{\chi}^0_a}^2,m_\phi^2)  \right. \nonumber\\
&&~~~~~~~~~~~~~~~~~~~~ \left.
     +  { Z_{ij} \over m_{\tilde{\nu}_i}^2-  m_{\tilde{\nu}_j}^2 }
         [B_0(m_{\tilde{\chi}^0_a}^2,m_{\tilde{\nu}_i}^2) -
          B_0(m_{\tilde{\chi}^0_a}^2,m_{\tilde{\nu}_j}^2) ] \,
   \right)  
\end{eqnarray}
where $N_{ab}$ is the 4x4 neutralino diagonalization matrix,
$\tilde{\chi}^0$ denotes the neutralino mass eigenstates, 
$\phi$ represents the neutral Higgs bosons   ($\phi= h, H$ and $A$),
and the loop-function $B_0$ is given by 
$B_0(x,y)=-{x\over x-y} \ln{x\over y}- \ln{x\over Q^2} +1$ 
with the renormalization scale $Q$.
The effect of the bilinear R-parity violating terms are encoded  in 
the coefficients $\theta_{i\phi}$  and $Z_{ij}$ which are given by
\begin{eqnarray} \label{thetas}
 \theta_{ih} &=& +\xi_i s_\alpha+ \eta_i s_\beta m_A^2 
    { m_{\tilde{\nu}_i}^2 c_{\alpha-\beta}
    - M_Z^2 c_{2\beta} c_{\alpha+\beta} \over 
      (m_{\tilde{\nu}_i}^2-m_h^2) (m_{\tilde{\nu}_i}^2-m_H^2) }
      \nonumber\\
 \theta_{iH} &=& -\xi_i c_\alpha+ \eta_i s_\beta m_A^2 
    { m_{\tilde{\nu}_i}^2  s_{\alpha-\beta}
    - M_Z^2  c_{2\beta} s_{\alpha+\beta}  \over
      (m_{\tilde{\nu}_i}^2-m_h^2) (m_{\tilde{\nu}_i}^2-m_H^2) }
         \nonumber\\
 \theta_{iA} &=& -i \xi_i s_\beta +
         i \eta_i s_\beta { m_A^2 \over m_A^2- m_{\tilde{\nu}_i}^2}  
                  \nonumber\\
  Z_{ij} &=& \eta_i\eta_j 
        m_A^4 M_Z^2 c_\beta^2 s_\beta^4
        \left[ {m_{\tilde{\nu}_i}^2 \over  F_S^i} + 
               {m_{\tilde{\nu}_j}^2  \over F_S^j} \right]  ,
\end{eqnarray}
where $\eta_i$ is defined in Eq.~(\ref{etas}),
$\alpha$ is the usual diagonalization angle of two CP even 
Higgs bosons, and
$ F_S^i \equiv (m_{\tilde{\nu}_i}^2-m_A^2) (m_{\tilde{\nu}_i}^2-m_h^2)
(m_{\tilde{\nu}_i}^2-m_H^2)$.
Recall that the angle $\alpha$ is defined by
$c_{2\alpha}=c_{2\beta} (m_A^2-M_Z^2)/(m_h^2-m_H^2)$ and 
$s_{2\alpha}=s_{2\beta} (m_A^2+M_Z^2)/(m_h^2-m_H^2)$.

A few remarks are in order: (i)
The coefficients $\theta_{i\phi}$ are the linear combinations of
$\theta^S_{ij}$'s defined  in Eq.~(9) of Ref.~\cite{cjkp}. 
They are related by the Higgs mass diagonalization.  
In Eq.~(\ref{thetas}), the quantity $\xi_i$  appears to include 
the effect of neutrino-neutralino mixing by $\epsilon_i$.  
This $\xi_i$ dependence 
can be easily understood if one goes to the basis where $\epsilon_i$ 
vanishes \cite{CK}.  
(ii)  
The same diagrams have been considered in Ref.~\cite{sacha2} using the
mass-insertion method which must yield the equivalent results to ours.
These diagrams involve  two mass-insertions which can be seen here 
as products of two induced R-parity odd $\nu-\phi -\chi^0 $ vertices, 
$\theta_{i\phi} \theta_{j\phi}$, and as individual sneutrino 
vertices, $Z_{ij}$, which is R-parity even. 
(iii) Among various contributions in $\theta_{i\phi} \theta_{j\phi}$, 
the term proportional to $\xi_i \xi_j$ can be absorbed into the tree-level
mass term giving a negligible effect.  The term proportional to
$\xi_i \eta_j$ is suppressed due to the similar reason discussed before,
but cannot be neglected completely.
(iv)
The term $Z_{ii}$ is nothing but the contribution due to the 
sneutrino--anti-sneutrino mass splitting induced by R-parity violation,
a l\'{a} Grossman-Haber \cite{GH}, and $Z_{ij}$ with $i \neq j$
comes from the effective sneutrino mixing vertices, 
$\nu_i - \tilde{\nu}^*_j-\chi^0$.  
(v) The terms with $Z_{ij}$ are proportional to 
$M_Z^2 c_\beta^2/m_{\tilde{\nu}_i}^2$,  and thus
give smaller contributions than the terms with $\eta_i \eta_j$ from 
$\theta_{i\phi} \theta_{j\phi}$ in a reasonable range of parameters.  
However, they can give a sizable effect in general.

Now, let us consider the other one-loop contributions and 
show that (\ref{Mgloop}) dominates over them in the 
case of the general non-universality.  Among various 
contributions, we take the well-known 
diagram with squark--quark exchange to be compared with (\ref{Mgloop}). 
Considering  the trilinear couplings induced from bottom 
quark Yukawa couplings $h_b$ such as $\lambda'_{i33} = 
\epsilon_i h_b$, one has
\begin{equation} \label{Pibb}
  \tilde{\Pi}^\nu_{ij} \approx {3 \over 8 \pi^2} {h_b^2 m_b^2 \mu t_\beta
         \over m^2_{\tilde{b}} } \epsilon_i \epsilon_j .
\end{equation}
Taking the ratio of the above two contributions, one typically gets 
$(\ref{Pibb})/(\ref{Mgloop}) \approx 5\times 10^{-6}\, t_\beta^3 \,
(\epsilon/\eta)^2$ with $m_{\tilde{\chi}^0}=100$ GeV, $\mu=m_{\tilde{b}} =
250$ GeV.  Therefore, (\ref{Pibb}) can be neglected as far as 
$\tan\beta$ is not too large and $\epsilon_i \sim\eta_i$.   
In the similar way, one can find that
the other diagrams are also sub-leading to (\ref{Mgloop}).
In Ref.~\cite{anjan1}, a slight deviation of non-universality has been 
assumed to yield $\epsilon/\eta \sim 10^{3}$ and thus (\ref{Pibb}) was 
considered as the main one-loop correction.  In fact, this is a 
typical situation in the case of universality.
The importance of the contribution (\ref{Mgloop}) in the case of 
large deviation from  universality has been notified 
in Ref.~\cite{sacha2} and its impact on viable neutrino mass matrices   
has been considered in Refs.~\cite{sacha1,abada}. 

{} From the previous discussions, we can write down the leading contributions
to the full mass matrix (\ref{Mn}) as follows:
\begin{equation} \label{Mfinal}
 M^\nu_{ij} \approx - {M_Z^2\over F_N} \xi_i\xi_j c_\beta^2
  - {g^2\over 32\pi^2}\sum_a m_{\tilde{\chi}^0_a} \eta_i \eta_j f_{ij}^a
\end{equation}
where $f^a_{ij}$ derivable from Eqs.~(\ref{Mgloop}) and (\ref{thetas})
is the function of the masses of neutralinos, 
sneutrinos and  Higgs bosons and its flavor dependence comes from the
non-universal slepton masses.

\medskip

We are ready to discuss how the desirable neutrino masses 
and mixing can be realized by the bilinear R-parity violation with 
generic non-universal soft masses.  For this,  we  will take 
the following representative set of R-parity conserving parameters;
\begin{equation} \label{xxx}
t_\beta=5,\quad m_A=300,\quad \mu=-250,\quad M_2=2M_1=200,
\end{equation}
throughout this paper.  This choice gives
the light and heavy neutral Higgs boson masses,
$m_h=84$ GeV and $m_H=302$ GeV, respectively.
Other choices will not change the main features of our results.
Concerning the R-parity violating parameters, we allow the
general flavor dependence for the supersymmetric $\epsilon_i$
and soft $B_i$ parameters.  To make our discussion simpler, we will
take $m^2_{L_i H_1}=0$ in this paper. 
This would be a plausible choice for the  minimal lepton flavor
violation as it may arise due to some mechanism of 
generating  the $\mu$ and $\epsilon_i\mu$ terms.

Now, let us start with the simplest case:
{\it (A)  the ``minimal'' deviation from the universality}, that is,
sleptons have a universal soft-mass:
$m_{H_1}^2\neq m^2_{L_1}=m^2_{L_2}= m^2_{L_3}$.
This  was the scheme employed in the analysis of Refs.~\cite{sacha1,sacha2}.
In this case, the lepton flavor dependence in $f_{ij}$ disappears
and thus the neutrino mass matrix (\ref{Mfinal}) takes
the following simple form:
\newcommand{\x}{{\rm x}}
\newcommand{\y}{{\rm y}}
\begin{equation} \label{Mmfv}
 M^\nu_{ij} \approx
  m_\x \hat{\x}_i \hat{\x}_j + m_\y \hat{\y}_i \hat{\y}_j
\end{equation}
where $m_\x=|\xi|^2M_Z^2 c_\beta^2/F_N$,
$m_\y \sim |\eta|^2  m_{\tilde{\chi}^0}g^2/64\pi^2$. Here, 
$\hat{\x}$ and $\hat{\y}$ are nothing but the unit vectors in
the direction of $\vec{\xi}$ and $\vec{\eta}$, respectively.
As analyzed in Ref.~\cite{hwang}, the mass matrix (\ref{Mmfv}) has two
non-vanishing eigenvalues, $m_3 \approx m_\x$ and 
$m_2 \approx m_\y s^2_\varphi$, whose eigenvectors are in the directions of
$\hat{\x}$ and $\hat{\x} \times (\hat{\x} \times \hat{\y})$, respectively.
Here the angle $\varphi$ is defined by 
$c_\varphi = \hat{\x}\cdot\hat{\y}$.  From these, 
one finds that the desired neutrino mixing matrix (\ref{mns}) is 
obtained for
\begin{eqnarray} \label{xandy}
&& \hat{\x} \simeq (\theta_{13},  {1\over\sqrt{2}}, {1\over\sqrt{2}}),
\quad\mbox{and} 
     \nonumber\\
&& \hat{\y} \propto (s_{12}, \sqrt{2}(1+k)  c_{12},  
                \sqrt{2} k c_{12}) \,,
\end{eqnarray}
with an arbitrary number $k$.
The ratio of two mass eigenvalues is  given by
\begin{equation} \label{m2overm3}
 {m_2 \over m_3} \sim {g^2 \over 32\pi^2} 
                 {m_{\tilde{\chi}^0} F_N \over M_Z^2}
                 t_\beta^2 {|\eta|^2 \over |\xi|^2} s^2_\varphi \,.
\end{equation}
Note that one can easily obtain its right value to
accommodate the atmospheric and solar neutrino (LMA) mass scales; namely,
$m_2/m_3 \approx \sqrt{\Delta m^2_{sol}/
\Delta m^2_{atm}} \sim 0.16$ putting
$m_{\tilde{\chi}^0}= F_N =200$  GeV, 
$t_\beta = 5$, $|\eta|/|\xi|=1$ and $s^2_\varphi=1$. 
Furthermore, the relation (\ref{xandy}) can also be arranged 
by an appropriate choice of two independent set of parameters 
$\xi_i$ and $\eta_j$.  In the similar way, 
the LOW solution can also be easily accommodated.
However, it remains to be seen how such 
an arrangement for $\xi_i$ and $\eta_i$ can be made 
in terms of the input parameters, 
$\epsilon_i$, $\Delta B_i/B$ and $\Delta m^2_i/m_{H_1}^2$.
In order to answer this question, let us 
choose the following set of values;
$$\xi_i =  (0.1,1,1), \qquad
   \eta_i \propto (\sqrt{2} t_{3}, 1, -1)  
              \quad\mbox{with}\quad t_{3}=0.65  $$
which give rise to the desired bi-large mixing of the atmospheric and solar
neutrino oscillations.  Note that the above choice corresponds to 
$c_\varphi=0$. The normalization of $\eta$ will be chosen to 
reproduce a right value of $\Delta m^2_{sol}/
\Delta m^2_{atm} \sim 2\times10^{-2}$.
Since we will calculate the ratios of neutrino
mass eigenvalues and mixing angles, we put, e.g.,  $\xi_2=\xi_3=1$.
In order to obtain the mass scale of $m_3=0.05$ eV, one can take an
overall rescaling of R-parity violating variables, 
$\xi$, $\eta$ and $\epsilon$, by factor of $5\times10^{-6}$.  
We now give three examples realizing the above
choice of $\xi_i$ and $\eta_i$ as follows.

\noindent
(A1) $\Delta m_i^2/m_{H_1}^2=0.7$: This  corresponds to the sneutrino mass,
$m_{\tilde{\nu}_i}=67$ GeV, and gives the neutrino mass matrix,
$$
M^\nu_{ij}=-2.12\, \xi_i \xi_j + 0.18\, \eta_i \eta_j. 
$$
Therefore, the choice of $\eta_i=(1.1, 1.2, -1.2)$  leads to the desired
results as follows;
\begin{equation} \label{setA1}
{\Delta m^2_{sol} \over \Delta m^2_{atm}}=0.03, 
  \quad U_{e3} = 0.08,  \quad
   \sin^22\theta_{atm}=0.99, \quad
   \sin^22\theta_{sol}=0.82  .
\end{equation}
Our choice of $\xi_i=(0.1,1,1)$ and the above $\eta_i$ is realized by
the following input parameters;
$\epsilon_i =(4.5, 1.1, -9.5)$ and 
$\Delta B_i/B =(0.22, 0.18, 0.23)$.

\noindent
(A2) $\Delta m_i^2/m_{H_1}^2=-1$: It gives rise to 
$m_{\tilde{\nu}_i}=228$ GeV and 
$$
M^\nu_{ij}=-2.12\, \xi_i \xi_j + 0.089\, \eta_i \eta_j. 
$$
Taking $\eta_i=(1.4,1.5,-1.5)$, we find  
\begin{equation} \label{set2}
{\Delta m^2_{sol} \over \Delta m^2_{atm}}=0.018, 
  \quad U_{e3} = 0.07,  \quad
   \sin^22\theta_{atm}=0.99, \quad
   \sin^22\theta_{sol}=0.83 .
\end{equation}
The corresponding input parameters are
$ \epsilon_i =(-4.2, -3.4, 5.9) $ and $ \Delta B_i/B =(-0.31, -0.15, -0.42)$. 

\noindent
(A3) $\Delta m_i^2/m_{H_1}^2=0.1$:  It leads to 
$m_{\tilde{\nu}_i}=146$ GeV and 
$$
M^\nu_{ij}=-2.12\, \xi_i \xi_j - 0.0022\, \eta_i \eta_j. 
$$
With the choice of  $\eta_i=(9.2, 10, -10)$, we get  
\begin{equation} \label{set3}
{\Delta m^2_{sol} \over \Delta m^2_{atm}}=0.02, 
  \quad U_{e3} = 0.02,  \quad
   \sin^22\theta_{atm}=0.98, \quad
   \sin^22\theta_{sol}=0.84 ,
\end{equation}
and the input parameters; $ \epsilon_i =(283,287,-334)$  and 
$  \Delta B_i/B =(0.032, 0.031, 0.033) $.

For the cases (A1) and (A2),  our general parameter scan showed that
the realistic neutrino masses and mixing can be obtained
within the range of input parameters:
$1\lesssim |\epsilon_i|\lesssim10$ and 
$0.1\lesssim|\Delta B_i/B|\lesssim 1$ leading to
$|\xi|, |\eta|\sim 1$.  From the above samples, one can see that
there need certain arrangements in the flavor structure of the input 
parameters realizing the required  mixing angles.
This would be the case in many class of models.
In our case, the smallness of  $|\xi_1|$ is arranged not by the smallness
of $|\epsilon_1|$ but by a partial cancellation between two terms:
$\Delta m_1^2 \approx -\Delta B_1 \mu t_\beta$ leading to 
$\Delta B_1/B \approx 0.22$ and $-0.31$ for (A1) and (A2), respectively.
This pattern arises also in more general cases as we will see shortly.
Since $|\epsilon_1|$ is not necessarily smaller than $|\epsilon_{2,3}|$,
it is favored to have $\Delta B_1/B \sim \Delta B_{2,3}/B$.
Thus, a vanishingly small $|U_{e3}|$ cannot be naturally realized 
in our scheme.
In the case (A3), the universality is maintained to a certain degree, 
As we can see, this requires $|\epsilon_i| \gg |\eta_i|, |\xi_i|$ and 
a strong correlation for the fine-tunned values of $|\Delta B_i/B| \ll 1$.  
In fact, this is a characteristic property of 
the universality case where the small deviation of 
$\Delta m_i^2/m_{H_1}^2$  and $ \Delta B_i/B$ arises 
due to RGE of soft parameters.  We excluded such cases in our analysis.

\medskip

Let us now relax the universality condition of the slepton and Higgs boson 
masses, which leads to the following form of the neutrino mass matrix;
\begin{equation} \label{Mgfv}
 M^\nu_{ij}= c_0 \xi_i \xi_j + c_{ij} \eta_i \eta_j \,,
\end{equation}
where $c_0=-2.12$ again with the choice of Eq.~(\ref{xxx}) and the flavor
dependence in $c_{ij}$ appears due to non-universal slepton masses. 
We first consider an interesting case where 
{\it (B) the flavor independence is assumed for $\epsilon_i$} to see
whether only non-universality in soft-parameters can be the  source
of the bi-large mixing.  As an example, we take

\noindent
(B1)
$\epsilon_i=1$,
$\Delta m_i^2/m_{H_1}^2 =(-3.3, -3.1, -4.3)$ and 
$\Delta B_i/B =(-1.0, -2.6, -3.5)$:\\
This gives us 
$\xi_i=(-0.047,1.25, 1.27)$, $\eta_i=(-1.05, -1.35, -2.23)$ and  thus
$$ c_{ij}=\pmatrix{ 0.41 & 0.48 & 0.15 \cr
                    0.48 & 0.57 & 0.18 \cr
                    0.15 & 0.18 & 0.046\cr }. $$
As a result, we get
\begin{equation} \label{set4}
{\Delta m^2_{sol} \over \Delta m^2_{atm}}=0.017, 
  \quad U_{e3} = 0.14,  \quad
   \sin^22\theta_{atm}=0.94, \quad
   \sin^22\theta_{sol}=0.73 . 
\end{equation}
Again, one needs a relation $\Delta m^2_1 \approx -\Delta B_1 \mu t_\beta$.
We find that this case (B) is not particularly fine-tunned compared to
the previous case (A) and can be a viable option.  

Finally, we consider {\it (C) the most general case}
where we take arbitrary values of the 9 input parameters, 
$\epsilon_i, \Delta m^2_i/m^2_{H_1}$ and $\Delta B_i/B$, whose sizes 
are however restricted within the range of ($0.1-10$).
In FIGs.~1 and 2, we present the scatter plot in term of 
$x_i=m_{L_i}^2/m_{H_1}^2$ and $p_i=\Delta B_i /B$ with $i=1$ and 2, 
respectively, which generate the desired neutrino masses and mixing.  
FIG.~1 shows that  a solution set are centered 
around the values of $x_1$ and $p_1$ for which the cancellation in 
$\xi_1$ happens as discussed before.   Anther solution set is allowed
around $x_1=3.4$ or $0.4$ for which the sneutrino mass is close to
the heavy or light Higgs mass, respectively.  In this region, the 
mixing elements (\ref{thetas}) and thus the coefficients $c_{ij}$
in Eq.~(\ref{Mgfv}) become large to enhance the one-loop contribution.
As a consequence, $U_{e3}$ can be arranged to be small without 
making $\xi_1$ small.  In FIG.~2, one sees that the points ($x_2,p_2$)
close to ($x_1,p_1$) are favored although those points 
FIG.~1 allowing the cancellation in $\xi_1$ are excluded as can be expected.
The plot in terms of ($x_3$, $p_3$) is also very similar to FIG.~2.
In FIGs.~1 and 2, we plotted  only the points
where the tree mass is three times larger than the loop mass.
Here, let us remark that the one-loop mass can be even larger than 
the tree mass. That is, it is possible that the one-loop 
contribution proportional to $\eta_i\eta_j$ is the main source 
of the atmospheric neutrino mass and mixing angle 
while the tree mass generates the solar neutrino mass 
and mixing angle.  Even though such cases of the loop dominance 
cannot be neglected, there is a much larger parameter space allowed
in the case of the tree dominance as one can expect.  This can be 
seen in FIGs.~3 and 4  which plotted all the allowed points 
in terms of the induced variables $\xi_i$ which determine the tree mass
matrix as in Eq.~(\ref{Mn}).   These two figures show that there appears
the pattern, $|\xi_1|\ll |\xi_2|\approx |\xi_3|$, which gives rise to
$\theta_{13}\ll1$  $s_{23}\approx c_{23} \approx 1/\sqrt{2}$
as shown in Eq.~(\ref{xandy}) for the tree-dominance case.


\medskip

To conclude,  we showed  how naturally the realistic neutrino mass 
matrix can arise from bilinear R-parity violation assuming  non-universal 
soft-terms.  When generic non-universality is allowed and $\tan\beta$
is not too large,  the neutrino mass matrix is dominated by two contributions;
the tree-level mass and the one-loop mass from the so-called 
Grossman-Haber diagrams arising due to the sneutrino--Higgs mixing.
This was checked by our numerical calculation taking the full one-loop
renormalized neutrino mass matrix.
In this scheme, the loop-to-tree mass ratio falls naturally
into the right range to generate the desired values for 
$\Delta m^2_{sol}/\Delta m^2_{atm}$.
Considering nine input parameters, 
$\epsilon_i$, $\Delta B_i$ and $\Delta m^2_{i}$,
we analyzed the parameter space accommodating  two large 
($\theta_{12}$ and $\theta_{23}$) and one small ($\theta_{13}$) mixing 
angles. Typically, the smallness of $\theta_{13}$ is realized by a
cancellation between the terms contributing to $\xi_1$.  This
was shown by  some examples and also by the scatter plot of FIG.~1.
Such an arrangement would not be a severe fine-tunning of 
input parameters.  However, our scheme cannot provide a natural reason
for vanishingly small $\theta_{13}$ if it turns out so.
We presented the results accommodating only the LMA solution,
but  the similar conclusion can be drawn also in the case of 
the LOW solution as can be inferred from our discussions.

\medskip
\noindent
{\bf Acknowledgment}: 
EJC was supported by the Korea Research Foundation Grant,
KRF-2002-015-CP0060, and DWJ by BK21 project by the 
Ministry of Education, Korea.

\newpage

\begin{figure}
\epsfig{file=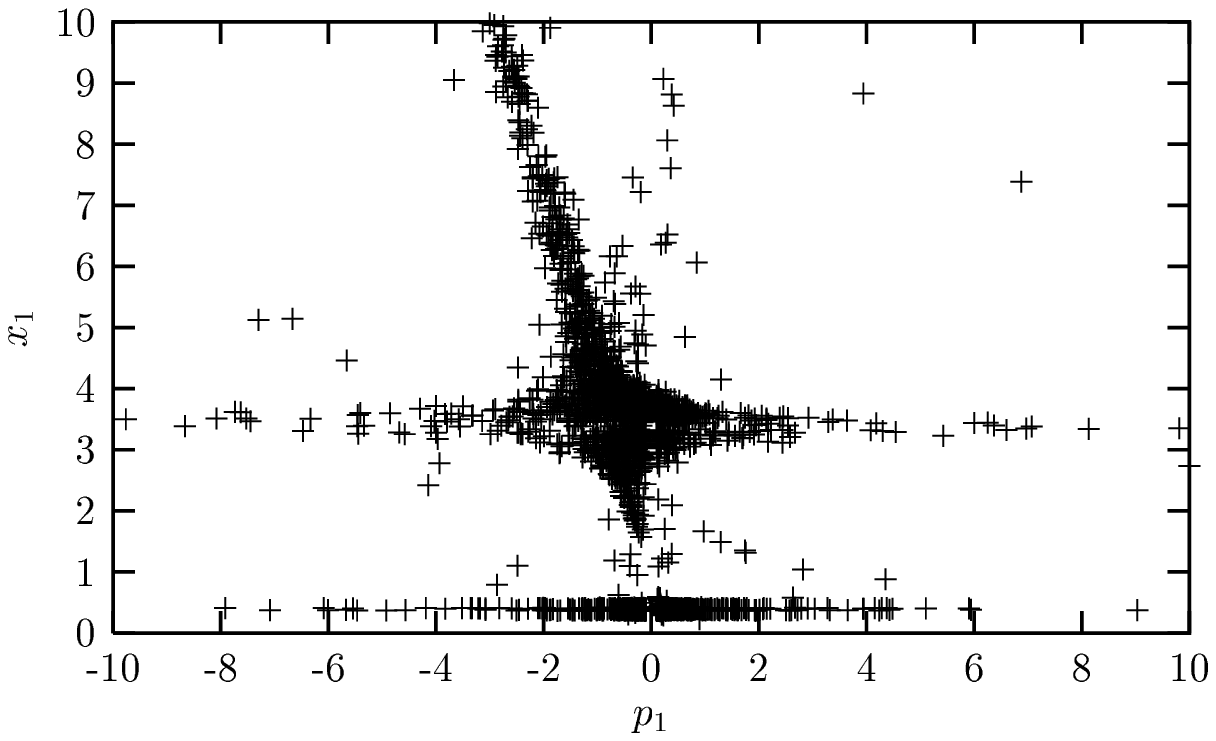,height=8cm,width=13cm}
\caption{ 
The tree-dominant points allowing the atmospheric and solar 
neutrino masses and mixing in terms of the two input variables,
$x_1=m_{L_1}^2/m_{H_1}^2$ and $p_1=\Delta B_1/B$.
}
\vspace{1.5cm}
\epsfig{file=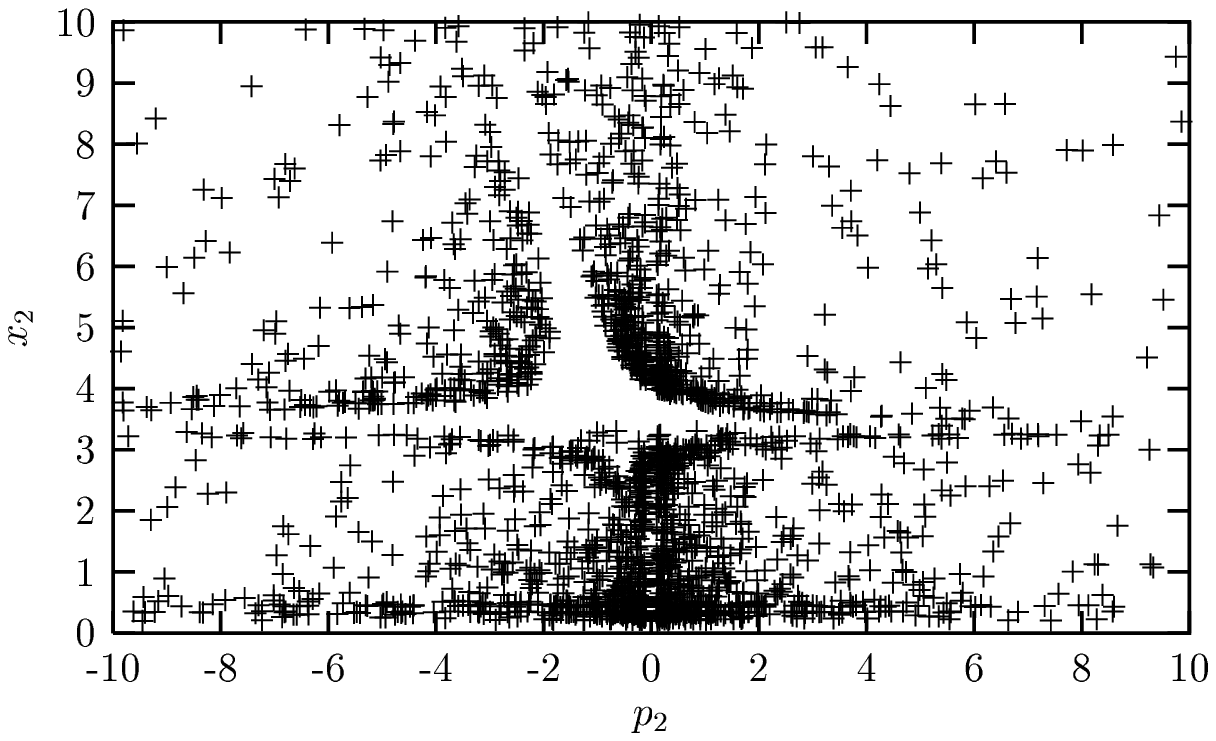,height=8cm,width=13cm}
\caption{ 
Same as FIG.~1 with $x_2=m_{L_2}^2/m_{H_1}^2$
and $p_2=\Delta B_1/B$.
}
\end{figure}

\begin{figure}
\epsfig{file=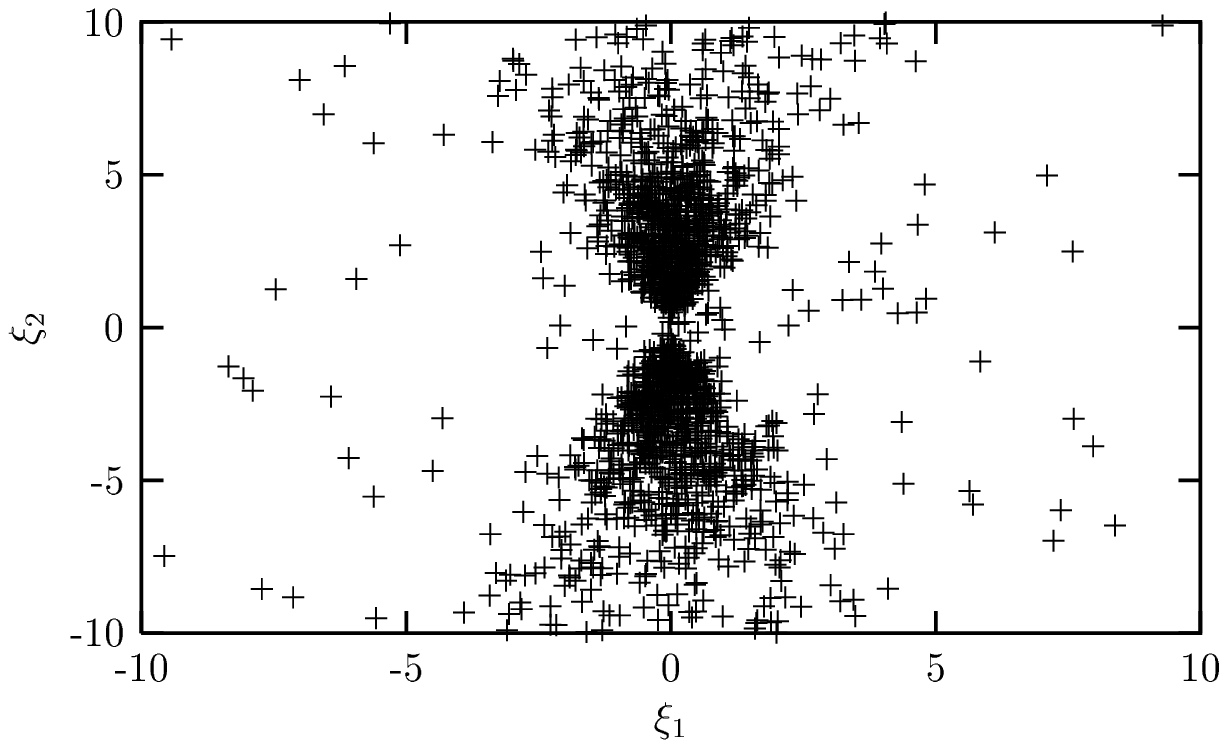,height=8cm,width=13cm}
\caption{ 
All the points allowing the atmospheric and solar neutrino 
masses and mixing in terms of the two induced variables, 
$\xi_1$ and $\xi_2$, controlling the tree-level mass matrix
as in Eq.~(\ref{Mn}).
}
\vspace{1.5cm}
\epsfig{file=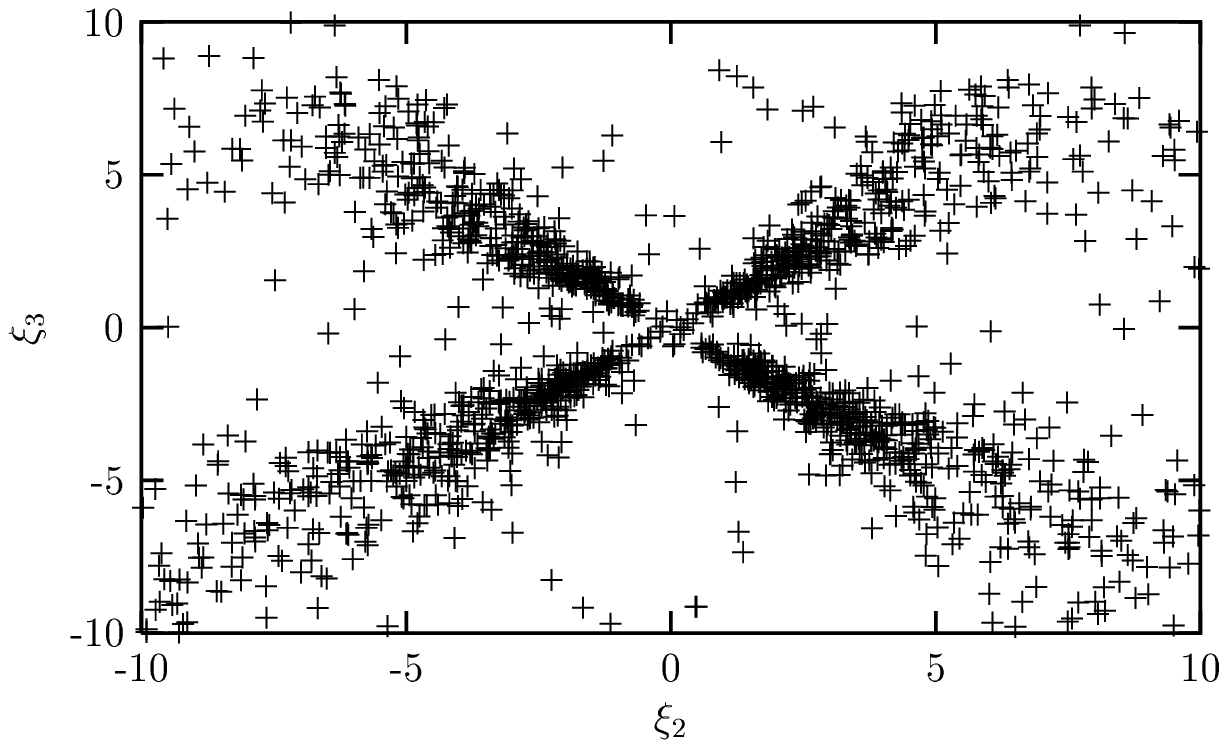,height=8cm,width=13cm}
\caption{ 
Same as in FIG.~3 with $\xi_2$ and $\xi_3$.
}
\end{figure}


\begin{thebibliography}{99}

\def\plb#1#2#3{Phys.\ Lett.\       {\bf B#1}  (#2) #3}
\def\npb#1#2#3{Nucl.\ Phys.\       {\bf B#1}  (#2) #3}
\def\prd#1#2#3{Phys.\ Rev.\        {\bf D#1}  (#2) #3}
\def\prl#1#2#3{Phys.\ Rev.\ Lett.\ {\bf #1}   (#2) #3}
\def\mpl#1#2#3{Mod.\ Phys.\ Lett.\ {\bf A#1}  (#2) #3}
\def\rep#1#2#3{Phys.\ Rep.\        {\bf #1}   (#2) #3}
\def\sci#1#2#3{Science             {\bf #1}   (#2) #3}
\def\astro#1#2#3{Astrophys.\ J.\   {\bf #1}   (#2) #3}
\def\epj#1#2#3{Eur.\ Phys.\ J.\   {\bf C#1}   (#2) #3}
\def\jhep#1#2#3{JHEP              {\bf #1}   (#2) #3}
\def\ptp#1#2#3{Prog.\ Theor.\ Phys.\ {\bf #1}  (#2) #3}

\bibitem{SK}
 Y. Fukuda {\it et. al.}, Super-K collaboration, \prl{81}{1998}{1562}.
\bibitem{SNO}
 Q.R. Ahmad {\it et. al.}, SNO collaboration, \prl{89}{2002}{011301}.
\bibitem{CHZ}
  CHOOZ collaboration, M. Apollonio {\it eta. al.}, \plb{420}{1998}{397}.
\bibitem{MNS}
 Z. Maki, M. Nakagawa and S. Sakata, \ptp{\bf 28}{1962}{870}.
\bibitem{HS}
 L. Hall and M. Suzuki, \npb{231}{1984}{419}.
\bibitem{CK}
 E.J. Chun and S.K. Kang, \prd{61}{2000}{075012}.
\bibitem{cjkp} 
 E.J. Chun, D.W. Jung, S.K. Kang and J.D. Park, \prd{66}{2002}{073003}.
\bibitem{valle1}
 M. Hirsh, M.A. Diaz, W. Porod, J.C. Romao and J.W.F Valle, 
 \prd{62}{2000}{113008}.
\bibitem{sacha1}
 A.\ Abada, S.\ Davidson and M.\ Losada \prd{65}{2002}{075010}.
\bibitem{anjan1}
 A.S. Joshipura and R.D. Vaidya, \npb{639}{2002}{290}.
\bibitem{abada}
 In light of recent SNO data, another possibility has been investigated 
 by  A. Abada, G. Bhattacharyya and M. Losada, \prd{66}{2002}{071701}.
\bibitem{fcnc}
  F. Gabbiani, E. Gabrielli, A. Masiero, and  L. Silvestrini,
  \npb{477}{1996}{321}.
\bibitem{jaja} B. Mukhopadhyaya, S. Roy and F. Vissani, \plb{443}{1998}{191};
  E.J. Chun and J.S. Lee, \prd{60}{1999}{075006}; S.Y. Choi {\it et. al.},
   \prd{60}{1999}{075002}; W. Porod {\it et. al.}, \prd{63}{2001}{115004}.
\bibitem{stau}  For the case of stau decay, see,
  M. Hirsch, W. Porod, J. Romao and J.W.F Valle, hep-ph/0207334;
  M.A. Diaz, R.A. Lineros and M.A. Rivera, hep-ph/0210182.
\bibitem{sacha2}
 S. Davidson and M.\ Rosada, \prd{65}{2002}{075025}.
\bibitem{GH}
 Y. Grossman and  H.E. Haber, \prd{59}{1999}{093008}.
\bibitem{hwang}
 K. Choi, E.J. Chun and K. Hwang, \prd{64}{2001}{033006}.


\end{thebibliography}
\end{document}